\documentclass[10pt, A4]{article}

\usepackage[a4paper, total={6in, 8in}]{geometry}
\usepackage[utf8]{inputenc}
\usepackage{authblk}
\usepackage{amsmath,amsfonts,amssymb}
\usepackage{graphicx}
\usepackage[T1]{fontenc}
\usepackage{siunitx}
\usepackage[euler]{textgreek}

\newcommand{\iu}{\mathrm{i}\mkern1mu}

\title{A Super-Condenser for Labelfree Nanoscopy}

\author[1,2,*]{Florian Ströhl}
\author[1]{Ida S. Opstad}
\author[1]{Jean-Claude Tinguely}
\author[1]{Firehun T. Dullo}
\author[2]{Ioanna Mela}
\author[2]{Johannes W.M. Osterrieth}
\author[1]{Balpreet S. Ahluwalia}
\author[2]{Clemens F. Kaminski}
\affil[1]{Department of Physics and Technology, UiT The Arctic University of Norway, NO-9037 Tromsø, Norway}
\affil[2]{Department of Chemical Engineering and Biotechnology, University of Cambridge, CB3 0AS Cambridge, UK}

\affil[*]{florian.strohl@uit.no}

\date{}

\begin{document}

\maketitle

\section*{Abstract}
Labelfree nanoscopy encompasses optical imaging with resolution in the 100~nm range using visible wavelengths.
Here, we present a labelfree nanoscopy method that combines Fourier ptychography with waveguide microscopy to realize a \textit{super-condenser} featuring maximally inclined coherent darkfield illumination with artificially stretched wave vectors due to large refractive indices of the employed Si$_3$N$_4$ waveguide material.
We produce the required coherent plane wave illumination for Fourier ptychography over imaging areas 400~\textmu m$^2$ in size via adiabatically tapered single-mode waveguides and tackle the overlap constraints of the Fourier ptychography phase retrieval algorithm two-fold: 
firstly, the directionality of the illumination wave vector is changed sequentially via a multiplexed input structure of the waveguide chip layout and secondly, the wave vector modulus is shortend via step-wise increases of the illumination light wavelength over the visible spectrum. 
We validate the method via in silico and in vitro experiments and provide details on the underlying image formation theory as well as the reconstruction algorithm. 

\section{Introduction}

Conventional nanoscopy, optical microscopy with resolution below 100~nm, is based on fluorescence\cite{schermelleh2019super}.
Often listed advantages of nanoscopy, especially in comparison to electron microscopy, are the simple sample preparation, live-cell compatibility, and molecular specificity.
Though live-cell compatible, the introduction of fluorescent labels onto the molecular structures of interest are in living cells likely to cause both functional and structural aberrations, potentially leading to false conclusions, and is also associated with problems like photobleaching and phototoxicity, variable label specificity, imaging- and image reconstruction-related artifacts, and lengthy optimization protocols \cite{van2011direct,demmerle2017strategic}.
The advantage of label specificity also has its downside of excluding (ultra)-structural context of the specifically labeled structure, although this can be alleviated to some degree via multi-channel labeling. 
Synergistic approaches combining the advantages of label specificity from conventional nanoscopy together with ultra-structural context obtained via label-free nanoscopy, could bring many new insights about cellular functions, especially as (contrary to correlative light and electron microscopy\cite{van2008correlative}), label-free (optical) nanoscopy has the potential of also being applied to living cells and cellular systems. 
Excluding methods that have not gone beyond the proof-of-principle stage like hyperlensing\cite{rho2010spherical} or super-oscillation microscopy\cite{kozawa2018superresolution}, suitable labelfree methods that have the potential to provide nanoscopic resolution can be sorted broadly into four groups:
(1) \textbf{Autofluorescence probed with conventional nanoscopy}.
Although certain critical fluorophore properties that are required for ultra-high resolution nanoscopy like photo-switching\cite{betzig2006imaging} are normally not present in intrinsically fluorescent samples, structured illumination microscopy has been shown to resolve features in the 150~nm regime in unlabeled retinal tissue\cite{best2011structured}.
Note that intrinsic fluorescence is a property present only in some but not all samples.
(2) \textbf{Nearfield scanning optical microscopy}\cite{betzig1992near}, which rasters a sample with an effective resolution below 100~nm. 
Akin to electron microscopy, this scanning optical approach has a low through-put and is challenging to combine with fluorescence-based nanoscopy.
(3) \textbf{Deep ultra-violet microscopy} - a theoretically simple approach as resolution scales linearly with employed imaging wavelength.
However, the limited availability and performance of optical components in this spectral range as well as the high phototoxicity associated with ultraviolet radiation offset the benefits gained by wavelengths below 400~nm illumination.
(4) \textbf{Fourier ptychography}, FP\cite{Zheng2013}, a technique specifically developed for improving digital pathology\cite{horstmeyer2015}.
In FP the sample is illuminated and imaged sequentially with plane waves from a multitude of directions that densely sample the illumination condenser numerical aperture ($NA_c$).
The generated set of images is then synthesized into a super-resolved amplitude and phase image of resolution $\Delta x$ given by

\begin{equation}
	\Delta x = \frac{\lambda}{NA_{c} + NA_{o}}
	\label{eq:abbe_incoherent_separate}
\end{equation}

\noindent via a dedicated \textit{phase retrieval algorithm}\cite{Zheng2013}.
It is well known that in conventional microscopy the condenser $NA_c$ used for illumination should be matched to (or even slightly below) the objective $NA_o$ \cite{abbe1873}, resulting in an effective maximal resolution of $\Delta x = \frac{\lambda}{2 NA_o}$, the Abbe resolution limit \cite{abbe1873}.
Crucially in FP, a numerical aperture $NA_c$ of the condenser \textit{can} be larger than the objective’s $NA_o$ in order to increase resolution with respect to the detection objective.

\section{Theory}

\begin{figure}[t]
    \centering
    \includegraphics[width=\linewidth]{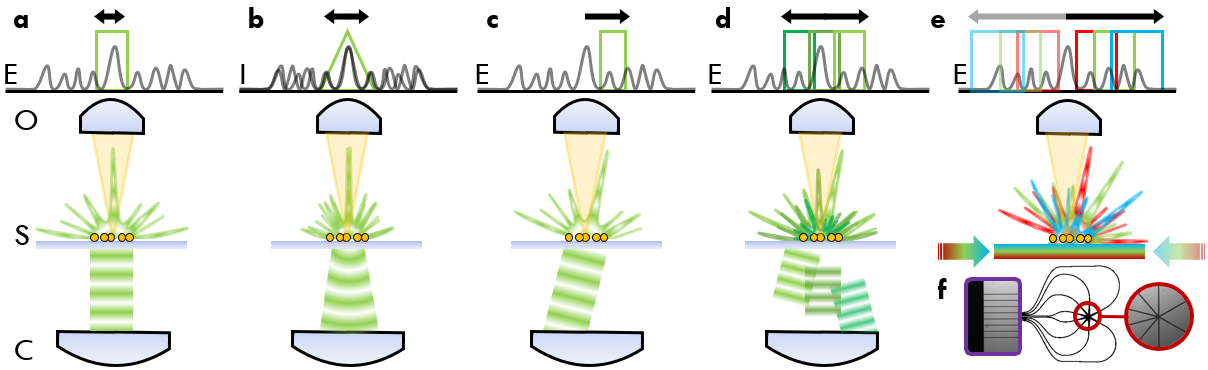}
    \caption{Amplitude/modulation transfer functions using (a) coherent, (b) incoherent, and (c) oblique illumination.
    Amplitude transfer function sampling in (d) conventional FP and (e) waveguide-based Fourier ptychographic microscopy.
    (f) The developed waveguide chip with multiple inputs.
    E: electric field, I: intensity, O: objective, S: sample, and C: condenser. 
    The arrow highlights the cut-off  frequency for different imaging modalities.}
    \label{fig:resolution}
\end{figure}

\noindent As the illumination in FP is coherent, the complex field (amplitude and phase) of the sample is probed rather than its intensity as in incoherent imaging.
Even though the effective aperture of a coherent microscope is half the size of an incoherent one's, plane wave illumination at oblique angles re-positions the sample's field in the aperture, thus giving access to finer details as visualized in Figure~\ref{fig:resolution}a-c.
Further, because the down-modulation of sample spatial frequency information with the illumination's spatial frequency occurs before being low-pass filtered by the objective aperture, access to information beyond $2NA_o$ is possible given high lateral spatial frequency of the illumination at greater angles than conventionally associated with $NA_o$.
To extract those finer details, multiple images are acquired sequentially using illumination angles spanning the entire condenser $NA_c$, and combined into a super-resolved image computationally (see Figure~\ref{fig:resolution}d).
Despite its potential, an extended condenser $NA_c$ has so far almost exclusively been used to increase the space-bandwidth product \cite{Zheng2013} rather than performing nanoscopy.
This is because, assuming the highest $NA$ available for both illumination and detection, the resolution caps at the incoherent resolution limit and is hence in the order of 200~nm.
To resolve nanoscopic structures using FP, a \textit{super-condenser} allowing illumination with spatial frequencies exceeding those offered by the best immersion objectives is necessary.
In the following, we show how such a super-condenser can be implemented in the form of a photonic waveguide chip in conjunction with multi-spectral illumination.
We show simulation results of the proposed method and perform proof-of-concept imaging of sub-diffraction-limit sized metal-organic framework (MOF) clusters.
Figure~\ref{fig:resolution}e outlines the fundamental mechanism of the proposed super-condenser, which aims to optimize the magnitude of the lateral illumination wave vector components and to provide as dense coverage of the virtual condenser pupil as possible.
To achieve largest lateral wave vector components, the illumination administered to the sample via photonic waveguides is intrinsically orthogonal to the detection objective and thus allows to maximize the wave vector magnitude geometrically.
Furthermore, akin to microscopy with immersion media, the illumination wave vector is stretched by a factor determined by the refractive index of the waveguide material.
Thus, by apt choice of material, a further tremendous increase in wave vector magnitude and image resolution can be achieved.
To illustrate, a conventional fluorescence microscope imaging GFP ($\lambda_{ex/em} = 488/512$~nm) with a high-performance 0.95~NA air objective offers a maximal resolution of $\lambda_{em}/2NA_o \approx 270$~nm.
The same sample imaged with the proposed super-condenser featuring Si$_3$N$_4$ waveguides with refractive index n $\approx 2.08$ yields a nominal resolution of $\lambda_{ex}/(\text{n}+NA_o) \approx 160$~nm.
Furthermore, it should be considered that incoherent microscopy techniques (like fluorescence or brightfield) have a strongly damped optical transfer efficiency for higher spatial frequencies, whereas the effective transfer function produced by Fourier ptychography has close to unity transmission strength, thus obtaining greatly enhanced contrast for finer structural details (as visualised in Figure \ref{fig:resolution}).

\section{Reconstruction Algorithm}

\begin{figure}[t]
    \centering
    \includegraphics[width=\linewidth]{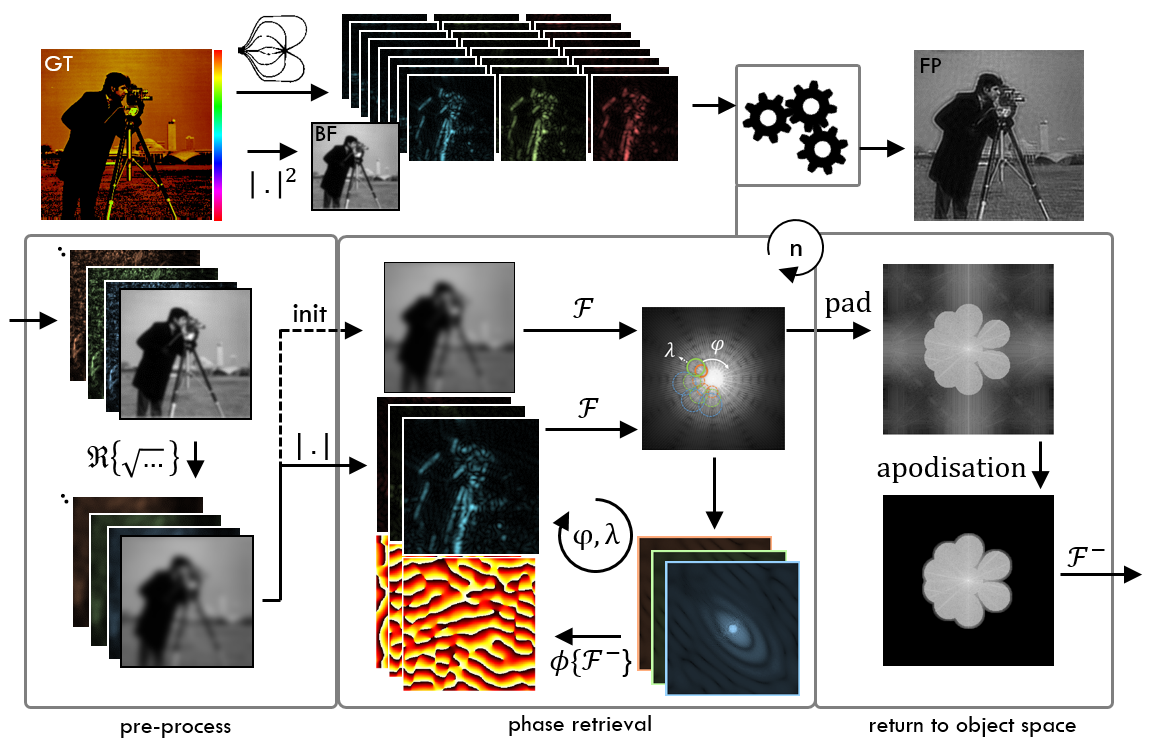}
    \caption{Phase-retrieval algorithm on simulated data. 
    Details are provided in the text.}
    \label{fig:algorithm}
\end{figure}

As our technique is based on Fourier ptychography it uses a phase retrieval algorithm\cite{Zheng2013}, which was slightly modified and is depicted in the box of Figure~\ref{fig:algorithm}.
The algorithm aims to invert the imaging pipeline and thus requires a detailed description of the image formation.
Let the raw images be denoted as $i_{k_0,k_c}(x)$, with $k_0$ being the illumination wave vectors’ lateral component, and $k_c$ the coherent cut-off frequency of the used wavelength, i.e. $k_c = \frac{NA_o}{\lambda}$. 
Using a coherent imaging model on the sample's complex field $s(x)$, which is illuminated with plane waves featuring wave vector $k_0$ and imaged by an objective characterized by the coherent point spread function $h_c(x)$, the coherent image formation equation reads 

\begin{equation}
    i_{k_0,k_c}(x) = \left| \left[ s(x) \times \text{exp}\left( -\iu k_0 \times x \right) \right] \otimes h_c(x) \right|^2.
\end{equation}

\noindent In this equation, $\iu$ is the imaginary unit, $\otimes$ is the convolution operator, and $h_c(x)$ is defined as a circle centred on the spatial frequency coordinate origin and with value one inside and zero outside its radius $k_c$.
The goal of the phase retrieval algorithm is to find the amplitude $a(x)$ and phase $\phi(x)$ component of the complex sample $s(x) = a(x) \times \text{exp}(\iu \phi(x))$. 
Three pre-processing steps are performed: (1) the raw data is background corrected, (2) then low-pass filtered, and (3) finally an initial guess of the amplitudes $a_{k_0,k_c}(x)$ is made. 
In analogy to the approach of Zheng\cite{Zheng2013} this is done by subtraction of a background estimate value $b$, and multiplication of a low-pass filter defined by the support of the incoherent optical transfer function to the image spectra $I_{k_0,k_c}(k)$ to remove noise from outside the pass-band of the objective, i.e. beyond the incoherent cut-off spatial frequency. 
As the incoherent cut-off frequency is twice the coherent cut-off frequency, a scaled version of the coherent transfer function can be used, i.e. $H_c\left( \frac{k}{2} \right)$. 
Note that Fourier analogues of real space functions, obtained via Fourier transform $\mathfrak{F}$ will be denoted via capitalization, so e.g. $H_c(k) = \mathfrak{F}\{h_c(x)\}$, with $k$ being the spatial frequency coordinate.
The inverse Fourier transform is written as $\mathfrak{F}^{-}$.

\noindent After low-pass filtering, the real part $\Re$ of the square root is taken to approximate the field distribution that formed the recorded intensities: 

\begin{equation}
    a_{k_0,k_c}(x) = \Re \left\{ \sqrt{ \mathfrak{F}^{-} \left\{ \mathfrak{F}\left\{i_{k_0,k_c}(x)-b\right\} \times H_c \left( \frac{k}{2} \right) \right\}}  \right\}.
\end{equation}

\noindent The phase retrieval part of the algorithm is then initialized using the estimated amplitude of a brightfield image as starting guess $f^0(x)$ \cite{dong2015incoherent} for the high-resolution Fourier ptychography image. 
In each iteration up to a total of $n$ rounds, $f^j(x)$ is sequentially updated for all available coherent illumination wave vectors $k_0$. 
The sequence is chosen such that the respective sub-sampled parts of Fourier space (which are centred around $k_0$ and with radius $k_c$), are \textit{spiralling out} from lower to higher spatial frequencies.
Formally in the algorithm, the individual updates are performed in three steps.
First, a temporary low-resolution image $t^j(x)$ is calculated from the Fourier ptychography estimate $f^j(x)$ for the current respective illumination featuring wave vector $k_0$, cut-off frequency $k_c$ and amplitude transfer function $H_c$ as

\begin{equation}
    t^j_{k_0,k_c}(x) = \mathfrak{F}^{-} \left\{ F^j(k-k_0) \times H_c \right\}.
\end{equation}

\noindent The phase $\Phi (t(x))$ of the temporary low-resolution image $t^j(x)$ is taken as an estimate of the phase distribution $\phi(x)$ of the sample $s(x)$. 
Hence, only the amplitude of $t^j(x)$ is updated, i.e. replaced by the estimated amplitude $a_{k_0,k_c}(x)$ of the respective pre-processed raw image

\begin{equation}
    t^{j+1}_{k_0,k_c}(x) = a_{k_0,k_c}(x) \times \text{exp} \left( \iu \Phi(t^j(x)) \right).
\end{equation}

\noindent The updated temporary image’s spectrum $T^{j+1}_{k_0,k_c}(k)$ is successively used to replace the respective region in Fourier space of the Fourier ptychography image's spectrum $F^{j+1}(k)$.
This region is centred on $k_0$ within a support area defined by the coherent transfer function $H_c(k)$ of that respective wavelength:

\begin{equation}
    F^{j+1}(k) = F^j(k) \times \left( 1-H_c(k-k_0)) \right) + H_c(k-k_0)) \times T^{j+1}_{k_0,k_c}(k-k_0).
\end{equation}

\noindent After each loop the lower spatial frequencies can be updated using an incoherent brightfield image in analogy to the updating step with evanescent illumination.
After $n$ full loops, the final Fourier ptychography image $f(x)$ is produced via apodization and a successive inverse Fourier transform with enlarged Fourier support (potentially made to fit via additional zero-padding) to yield a smoother transform result

\begin{equation}
    f(x) = \mathfrak{F}^{-} \left\{ apo(pad(F(k))) \right\}.
\end{equation}

\noindent For comparison to brightfield data, an intensity image can be created via squaring of the amplitude part of the Fourier ptychography reconstruction.
The presented algorithm was tested on simulated data, as displayed in the top of Figure~\ref{fig:algorithm} for a ground truth (GT) input and a successful FP intensity reconstruction.

\section{Experimental Results}

\begin{figure}[t]
    \centering
    \includegraphics[width=\linewidth]{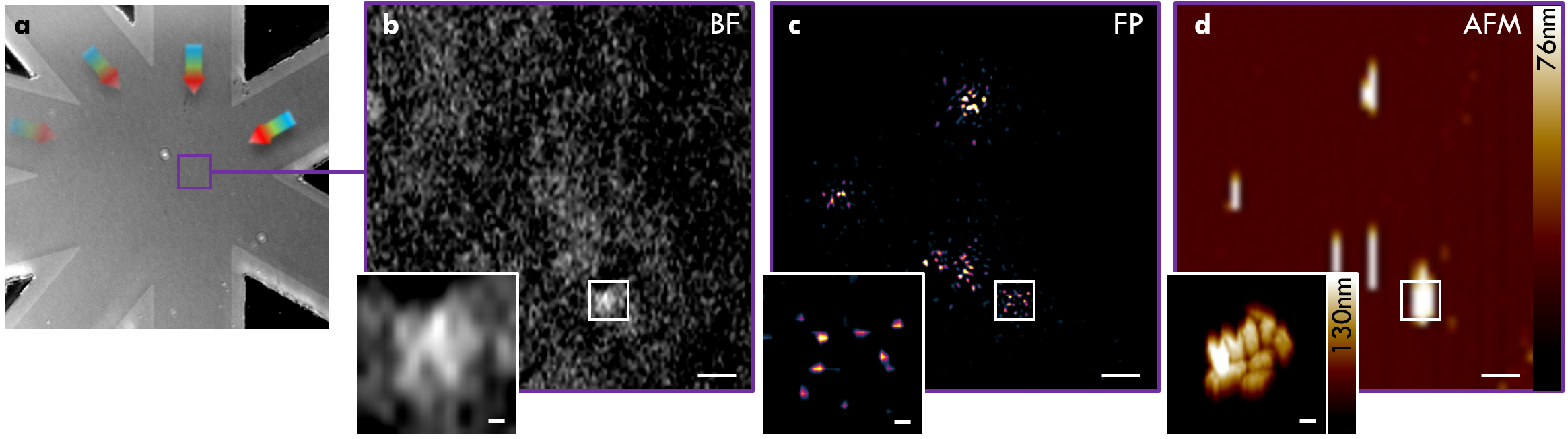}
    \caption{Imaging of metal-organic frameworks (MOFs).
    (a) Overview of the imaged region. 
    (b) Brightfield image using 490~nm LED light. 
    (c) Intensity image created by Fourier ptychography.
    (d) Atomic force microscopy image (line levelling artefacts prohibit a clear view of individual particles). 
    Inlays show a zoomed region of a cluster of MOFs.
    The overview image (a) measures 100$\times$100~\textmu m$^2$ and the scalebars in (c-e) are  1~\textmu m and 100~nm in the inlays respectively.}
    \label{fig:experiments}
\end{figure}

To validate our approach experimentally, we imaged clusters of metal-organic frameworks (MOFs).
The imaged MOFs belong to the group of Zirconium-MOFs with gold nano-rod core and have a small size distribution centered around 200~nm\cite{osterrieth2019core}.
To ensure adherence of the MOFs to the waveguides, the waveguide chip was plasma-treated for 40~s at 40~W using a 0.35~mbar oxygen atmosphere.
Then, a highly diluted aqueous solution of MOFs was drop-casted onto the waveguide chip imaging area and left to dry under a slight angle to provide a more even distribution of the particles.
After imaging with the super-condenser, a ground truth image of the sample was generated via atomic force microscopy (AFM) using a commercial system (Bioscope RESOLVE, Bruker).
The AFM was operated in tapping mode and RTESPA probes (Bruker) with a nominal spring constant of 6~N/m and resonant frequency 150~kHz.
A line scanning resolution of 256 lines with 256 samples/line for 50$\times$50~\textmu m$^2$ was used to generate an overview and 1168 lines with 1168 samples/line for 20$\times$20~\textmu m$^2$ were used for greater detail of selected areas.
To counter drift between frames, the individual raw frames were further aligned to each other using semi-automated alignment via the image processing software \textit{line ROI image alignment} in Fiji\cite{schindelin2012fiji}.
As shown in Figure~\ref{fig:experiments}, we retrieved images displaying both enhanced contrast and features beyond the incoherent Abbe diffraction limit.

In panel (a), an overview of the waveguide chip geometry is shown using a brightfield reflectance image (LED illumination centred at 490~nm). 
As shown in (b), the sample is only barely visible in a 490~nm brightfield incoherent illumination image (for the same objective), which displays an
unresolved cluster of 370~nm size in terms of full width at half maximum of a Gaussian fit (data not shown).
Displayed in panel (d), the underlying distribution of individual MOFs in this cluster is shown using atomic force microscopy (AFM) and finest scanning reveals dense clustering of particles.
The same distribution of clusters that is visible in the AFM image is also present in the Fourier ptychography image in panel (c), which potentially even resolves individual gold nanorod cores in the clusters.

\section{Discussion and Conclusion}

Care must be taken in interpreting these results as the discernible spots produced by the FP phase retrieval algorithm cannot be associated to individual particles in the AFM recordings with absolute certainty.
For instance, although the clusters visible in FP can be mapped onto the AFM ground truth, the individual particles are slightly misplaced or even missing completely.
This can have multiple reasons. 
Firstly, the displacement might be traced back to drift of the sample during raw data acquisition. 
A shift of sample information even in the nanometer range might thus produce artefacts causing erroneous particle localization.
Although drift was compensated for computationally, it cannot be avoided completely.
Secondly, the gold core of the MOFs is asymmetric and cause scattering preferentially in certain directions. 
Paired with variable signal strength for different illuminations caused by higher losses of the waveguides in longer arms, this could lead to some particles \textit{outshining} others in the reconstruction.
Also note that the amplitude of the scattered light is wavelength dependent.
Despite trying to adjust for this via variation of the laser illumination intensity and the exposure time of the camera, a limited maximum number of photons in certain raw frames restricted the achievable signal-to-noise ratio (SNR). 
Residual wavelength-dependent waveguide autofluorescence increases the challenge of limited SNR additionally.
In the future, more advanced algorithms originally developed for conventional Fourier ptychography could be adopted to computationally alleviate some of these concerns\cite{bian2013adaptive}.
Furthermore, axial chromatic offset could not be accounted for, which might stem from residual imperfections of the employed apochromatic objective and achromatic tube lens.
This might be tackled via fine z-stepping of the objective (not possible in the presented set-up which offers no finer than 0.5~\textmu m steps) and post-acquisition alignment or pre-acquisition calibration (potentially using multiple cameras to increase acquisition speeds).
Still, the produced images are promising for the young field of labelfree chip-nanoscopy.

\begin{figure}[t]
\centering
\includegraphics[width=0.5\linewidth]{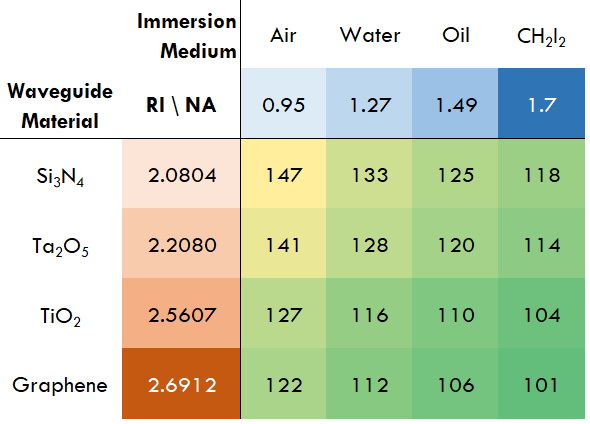}
\caption{Theoretically achievable resolution given in nm via different waveguide materials and substrate/immersion objective combinations (assuming shortest illumination wavelength of 445~nm).}
\label{tab:1}
\end{figure}

Looking ahead, a multitude of further developments might appear in close reach, which are in parts also conceptually simple.
An example would be a change of the chip material.
The catch here is the unavailability of established production methods for single-mode waveguides of more exotic character.
Nevertheless, promising candidates for alternative waveguide materials that also keep propagation losses low at short wavelengths are Ta$_2$O$_5$ \cite{gao2012exploitation}, TiO$_2$ \cite{siefke2016materials}, or graphene \cite{weber2010optical}) - see Figure \ref{tab:1}.
TiO$_2$ is an especially interesting material, as it transmits even in parts of the near ultra-violet range, which would allow sub-100nm resolution:

\begin{equation}
    \Delta x_{TiO} = \frac{405~\text{nm}}{(1.49+2.66)} = 97.6~\text{nm}. 
\end{equation}

\noindent A further current bottleneck, the limited field of view, could be alleviated through use of a \textit{slab} region at the imaging area illuminated by un-tapered single-mode waveguides. 
Although initially propagating as \textit{circular} waves (the two-dimensional analog of spherical waves produced by free-space point sources), the waves emanating from the waveguide outlets would  be sufficiently close to plane waves already after a travelled distance of few wavelengths and thus suitable for quasi-coherent imaging.
For a further field of view enhancement, illumination from a limited number of directions (potentially even only single-sided) could be feasible for amplitude-only samples\cite{zhou2018analysis} and would simplify both waveguide geometry and deliverable powers to the imaging area due to reduced bending losses. 
More speculatively, on-chip lasers\cite{xie2017chip} are an option to simplify the microscopy set-up and circumvent coupling losses.
An intriguing alternative could further be the use of a broadly emitting fluorescent film to generate the illumination light with successive narrow-band emission filtering as proposed by Pang et al\cite{pang2019on}.
Although the illumination in this set-up is not coherent, FP algorithms have been developed that can be applied\cite{dong2015incoherent}.
To tackle drift of the sample during image acquisition, a fully automated coupling procedure might speed up slow manual coupling of various wavelengths and to different inputs.
A possible future improvement in this respect is input-multiplexing which is feasible via conventional fiber-array adaptors.
These are standard in the telecommunication industry but are to date mostly available for infrared wavelengths.

In conclusion, we have developed a labelfree nanoscopy method that combines Fourier ptychography with waveguide microscopy to realize a super-condenser.
The waveguide geometry allows the use of maximally inclined coherent darkfield illumination and additionally makes use of the large refractive index of Si$_3$N$_4$ as waveguide material to further double the illumination wave vector amplitudes as compared to air.
We demonstrated our method in silico and validated it in experimental imaging of metal organic frameworks that contained gold nano-rod cores.
As proven by atomic force microscopy, we were able to image MOF clusters successfully and could even infer the distribution of individual particles within the clusters.
Taken together, Fourier ptychography in combination with enlarged illumination wave vectors is a promising avenue to enable nanoscopic imaging without the requirement of extrinsic labels.

\section*{Funding}

FS acknowledges funding from the European Molecular Biology Organisation (\#7411) and Marie Skłodowska-Curie actions (\#836355).
BSA acknowledges funding from the European Research Council (\#336716).
CFK acknowledges funding from the Physical Sciences Research Council (EP/H018301/1), Medical Research Council (MR/K015850/1 and MR/K02292X/1), Wellcome Trust (089703/Z/09/Z), and Infinitus China Ltd.

\section*{Author contributions statement}

FS and BSA conceived the idea of waveguide-chip based label-free nanoscopy using Fourier ptychography.
FS further expanded the idea towards the use of different wavelengths to fill Fourier space, performed simulations and analyzed the data. 
JCT and FTD designed, produced, and characterized the waveguides. 
ISO prepared samples. 
FS and ISO built the chip-microscope and performed waveguide imaging. 
IM performed AFM measurements.
JWMO provided metal-organic frameworks.
FS wrote the first version of the manuscript and all authors contributed towards the writing of the manuscript. 
CFK and BSA provided research tools and supported the project. 

\section*{Disclosures}

BSA has applied for patent GB1705660.7 on an \textit{Optical component for generating a periodic light pattern} that covers the designs of the waveguide chip used in this work.
The other authors declare no competing interests.

% References
\bibliography{references}
\bibliographystyle{unsrt}

\end{document}